\documentclass[pdftex,smallcondensed,epjc3]{svjour3}          

\RequirePackage[T1]{fontenc}

\smartqed  

\RequirePackage{graphicx}
\RequirePackage{times}      
\RequirePackage{flushend}
\RequirePackage[numbers,sort&compress]{natbib}
\RequirePackage[colorlinks,citecolor=blue,urlcolor=blue,linkcolor=blue]{hyperref}

\usepackage{epsfig}
\usepackage{graphics}
\usepackage{bm}
\usepackage{color}
\usepackage{dcolumn}   
\usepackage{bm}     
\usepackage{bbm}       
\usepackage{amssymb}  
\usepackage{amsmath}
\usepackage{latexsym}
\usepackage{float}
\usepackage{ifthen}
\usepackage{caption,subfig}
\usepackage{enumerate}
\usepackage{url}
\usepackage{caption,subfig}
\usepackage{amsopn}

\usepackage{amsfonts}
\usepackage{multirow}
\usepackage{array}
\usepackage{booktabs}
\usepackage{rotating}

\usepackage{ulem}
\def\clap#1{\hbox to 0pt{\hss#1\hss}}

\usepackage{dcolumn}   
\usepackage[utf8]{inputenc}
\usepackage[dvipsnames]{xcolor}

\newcommand{\dd}{{\rm d}}

\newcommand{\Lag}{\mathcal{L}}

\newcommand{\mS}{\mathcal{S}}
\newcommand{\mT}{\mathcal{T}}
\newcommand{\mH}{\mathcal{H}}

\newcommand{\GB}{\mathcal{G}}

\newcommand{\be}{\begin{equation}}
\newcommand{\ee}{\end{equation}}
\newcommand{\bea}{\begin{eqnarray}}
\newcommand{\eea}{\end{eqnarray}}

\newcommand{\Gammab}{\bar{\Gamma}}
\newcommand{\nablab}{\bar{\nabla}}
\newcommand{\Rb}{\bar{R}}
\newcommand{\Gb}{\bar{G}}

\renewcommand{\bf}[1]{{\textbf{#1}}}
\renewcommand{\tt}[1]{\textit{#1}}

\newcommand{\mU}{\ensuremath\mathcal{U}}

\newcommand{\mpl}{M_{\rm Pl}}

\newcommand{\gt}{\tilde{g}}
\newcommand{\Rt}{\tilde{R}}

\journalname{Eur. Phys. J. C}

\begin{document}

\title{Revisiting the Stability of Quadratic Poincar\'e Gauge Gravity}

\author{Jose Beltr\'an Jim\'enez\thanksref{e1,addr1}\and Francisco Jos\'e Maldonado Torralba\thanksref{e2,addr2,addr3}}

\thankstext{e1}{e-mail: jose.beltran@usal.es}
\thankstext{e2}{e-mail: f.j.maldonado.torralba@rug.nl}

\institute{Departamento de F\'isica Fundamental and IUFFyM, Universidad de Salamanca, E-37008 Salamanca, Spain.\label{addr1}
\and
Cosmology and Gravity Group, Department of Mathematics and Applied Mathematics, University of Cape Town, Rondebosch 7701, Cape Town, South Africa.\label{addr2}
\and
Van Swinderen Institute, University of Groningen, 9747 AG Groningen, The Netherlands.\label{addr3}
}

\date{Received: date / Accepted: date}

\maketitle

\begin{abstract}
Poincar\'e gauge theories provide an approach to gravity based on the gauging of the Poincar\'e group, whose homogeneous part generates curvature while the translational sector gives rise to torsion. In this note we revisit the stability of the widely studied quadratic theories within this framework. We analyse the presence of ghosts without fixing any background by obtaining the relevant interactions in an exact post-Riemannian expansion. We find that the axial sector of the theory exhibits ghostly couplings to the graviton sector that render the theory unstable. Remarkably, imposing the absence of these pathological couplings results in a theory where either the axial sector or the torsion trace becomes a ghost. We conclude that imposing ghost-freedom generically leads to a non-dynamical torsion. We analyse however two special choices of parameters that allow a dynamical scalar in the torsion and obtain the corresponding effective action where the dynamics of the scalar is apparent. These special cases are shown to be equivalent to a generalised Brans-Dicke theory and a Holst Lagrangian with a dynamical Barbero-Immirzi pseudoscalar field respectively. The two sectors can co-exist giving a bi-scalar theory.  Finally, we discuss how the ghost nature of the vector sector can be avoided by including additional dimension four operators.
\end{abstract}

\section{Introduction}

One of the most appealing approaches to gravity is the use of a gauge guiding principle to construct theories in a somewhat analogous manner to the celebrated Yang-Mills theories. As first proposed by Sciama and Kibble~\cite{sciama1962analogy,Kibble:1961ba}, a natural group for this task is Poincar\'e, conformed by the homogeneous Lorentz  group $SO(3,1)$ supplemented with spacetime translations. This inhomogeneous piece is then responsible for a number of interesting consequences, one of which is the appearance of torsion as the field strength of translations, while the Riemann curvature is associated to the homogeneous part. For an extensive review of these theories see e.g. \cite{Gauge,Gauge2,Obukhov:2018bmf}. Another interesting gauge approach to gravity is based on the observation that the translations are non-linearly realised so one can use the standard coset construction applied to the translations regarded as the quotient $ISO(3,1)/SO(3,1)$. See for instance \cite{Delacretaz:2014oxa} for interesting applications of this idea to gravitational systems. 

Since the inception of the Poincar\'e Gauge Theories (PGTs), the different attempts to generalise the properties and theorems of General Relativity (GR) have been a very active field. Paradigmatic examples include the study of singularities~\cite{Stewart:1973ux,Trautman:1973wy,Cembranos:2016xqx,Cembranos:2019mcb,delaCruz-Dombriz:2018aal}, the Birkhoff theorem \cite{Neville:1979fk,Rauch:1981tva,delaCruz-Dombriz:2018vzn}, existence of exact solutions \cite{Bakler:1984cq,Obukhov:1987tz,Blagojevic:2015zma,Cembranos:2016gdt,Obukhov:2019fti,Ziaie:2019dmq}, cosmology \cite{Kerlick:1975tr,Yo:2006qs,Shie:2008ms,Chen:2009at,Baekler:2010fr,Ho:2015ulu}, the motion of particles~\cite{Hehl:2013qga,Cembranos:2018ipn} and stability analysis~\cite{Sezgin:1979zf,Sezgin:1981xs,Chern:1992um,Yo:1999ex,Yo:2001sy,Karananas:2014pxa,Vasilev:2017twr}. Unveiling the fields content along with their stable/unstable nature is of course one of the most important questions for the viability of the theories with a crucial impact on the reliability of their phenomenology. The main goal of this work will be to provide an exhaustive analysis of the PGTs from a different approach to the existing studies in the literature as well as giving a complementary and comprehensive understanding of their properties. As opposed to the perhaps more extensively used Lorentz bundle approach resorting to vierbein fields, we will work directly in the metric formulation. Thus, the fundamental fields will be the metric tensor $g_{\mu\nu}$, with 10 components in four dimensions, and the 24 components of the torsion that can be decomposed as
\be
T^\alpha{}_{\mu\nu}=\frac{2}{3}T_{[\mu}\delta^\alpha_{\nu]}+\frac16\varepsilon^{\alpha}{}_{\mu\nu\rho}S^{\rho}+q^\alpha{}_{\mu\nu},
\label{eq:torsiondec}
\ee
with $T_\mu=T^\alpha{}_{\mu\alpha}$, $S_\mu=\varepsilon_{\mu\alpha\beta\gamma} T^{\alpha\beta\gamma}$ and $q^\alpha{}_{\mu\nu}$ the trace, the axial part\footnote{The axial part is described by an axial vector in 4 dimensions, while in arbitrary dimensions $d$ it is given in terms of a $(d-3)$-form. Since this is dual to a $3$-form in $d$ dimensions, the axial part can always be described by means of a 3-form.} and the tensor component of the torsion that satisfies $q^\alpha{}_{\alpha\mu}=q_{[\alpha\mu\nu]}=0$ respectively. These three pieces are irreducible under the Lorentz group as real representations and correspond to $(\frac12,\frac12)$, $(\frac12,\frac12)$ and $(\frac32,\frac12)\oplus(\frac12,\frac32)$ respectively \cite{Hayashi:1979wj}. The properties and stability of these different fields will depend on the particular action describing their dynamics. In this respect, it is common to restrict to the so-called quadratic PGTs whose action is composed by the most general expression up to quadratic terms in the field strengths, i.e., the curvature and the torsion. The motivation comes from usual Yang-Mills theories that are also constructed with quadratic terms in the field strengths of the gauge fields. Most studies on the stability in this theory have been performed by analysing the perturbative spectrum on a Minkowski background \cite{Sezgin:1979zf,Sezgin:1981xs,Karananas:2014pxa,Vasilev:2017twr}. These analyses concluded that the corresponding spectrum contains one massless spin-2, two massive spin-2, two massive spin-1 and two spin-0 fields. Already in \cite{Sezgin:1979zf} it was shown that, in a Minkowski spacetime, all of these components  cannot propagate simultaneously without incurring pathologies. In particular, it was proven that the absence of ghosts and tachyons restrains the spectrum to contain at most three propagating components, among other restrictions on the parameters of the theory. The authors in \cite{Yo:1999ex,Yo:2001sy} performed a more complete hamiltonian analysis of PGTs (see also a more recent analysis in \cite{Blagojevic:2018dpz}) signalling that the introduction of non-linearities would impose further more stringent constraints. Moreover, they showed that the only modes that could propagate were two scalars with different parity. We shall arrive at the same conclusion from a different path.

Irrespectively of the particular action describing the theory, the general spectrum of PGTs looks worrisome because of the presence of the two additional massive spin-2 fields that are presumably going to interact non-trivially among themselves and with the graviton. Given the delicate structure of the allowed unitary interactions  for multiple spin-2 fields \cite{Hinterbichler:2012cn}, one may expect the appearance of Boulware-Deser ghosts \cite{Boulware:1973my} in the full theory, unless much care is taken in constructing the interactions. Furthermore, the massive spin-1 fields can also cause some ghost-like instabilities, except if the derivative interactions guarantee the non-dynamical nature of the temporal components as to avoid the associated Ostrogradski instabilities. Although the spin-1 sector is less prone to pathologies, as well as more flexible regarding the construction of interacting theories than the spin-2 sector, we will find that the absence of ghosts in the spin-1 sector in turn suffices to dramatically reduce the parameter space for stable PGTs. As a matter of fact, we will see how GR stands out once more as one of the very few non-pathological theories. We will only find an additional class of non-pathological theories that secretly describes a generalised Brans-Dicke theory and the Holst formulation of GR where the Barbero-Immirzi parameter is promoted to a pseudo-scalar field.

In order to analyse the stability of the quadratic PGTs in full generality, we will follow a procedure without fixing any background and without having to perform a detailed hamiltonian analysis. More specifically, this approach consists in performing an exact post-Riemannian expansion that unveils all the interactions among the different sectors in a recognisable form. Knowing the pathological origin of the interactions, mainly the non-minimal couplings, will then permit us to easily pinpoint the problematic terms in the action that will jeopardise the stability by introducing Ostrogradski ghosts. This analysis will show in a very transparent manner the general pathological character of the quadratic PGTs and the origin of the instabilities. We will then show how to suitably choose the parameters to obtain viable theories. Furthermore,  we shall argue how to construct PG Lagrangians that allow a healthy propagation of the vectors while avoiding ghost instabilities by including additional dimension 4 operators. \\

\bf {Conventions:} The metric signature is $(-+++)$ and the Riemann tensor is defined as $R_{\mu\nu\beta}{}^\alpha=2\partial_{[\nu}\Gamma^\alpha_{\mu]\beta}+\cdots$, while the Ricci tensor is $R_{\mu\nu}=R_{\mu\alpha\nu}{}^\alpha$. We will denote objects associated to the Levi-Civita connection with a bar while standard notation will be used for objects associated to the full connection. Symmetrisation and antisymmetrisation are defined with the customary normalisation factors. The torsion will then be $T^\alpha{}_{\mu\nu}=2\Gamma^\alpha{}_{[\mu\nu]}=\Gamma^\alpha{}_{\mu\nu}-\Gamma^\alpha{}_{\nu\mu}$.

\section{Quadratic Poincar\'e gauge theories}

\subsection{The theory}
The parity preserving PGTs under consideration in this work are described by the Lagrangian
\begin{align}
\mathcal{L}_{\rm PG}=&\frac12\mpl^2\Big({R}+a_{1}T_{\mu\nu\rho}T^{\mu\nu\rho}
+a_{2}T_{\mu\nu\rho}T^{\nu\rho\mu}+a_{3}T_\mu T^\mu\Big)+b_{1}{R}^{2}+b_{2}{R}_{\mu\nu\rho\sigma}{R}^{\mu\nu\rho\sigma}\nonumber\\
&+b_{3}{R}_{\mu\nu\rho\sigma}{R}^{\rho\sigma\mu\nu}+b_{4}{R}_{\mu\nu\rho\sigma}{R}^{\mu\rho\nu\sigma}+b_{5}{R}_{\mu\nu}{R}^{\mu\nu}+b_{6}{R}_{\mu\nu}{R}^{\nu\mu},
\label{eq:PGaction}
\end{align}
with $\mpl^2$ the Planck mass and $a_i$ and $b_i$ some dimensionless parameters. Our approach will consist in performing an exact post-Riemannian expansion where the connection is decomposed into the Levi-Civita part of the spacetime metric $\Gammab^\alpha{}_{\mu\nu}(g)$ plus the contorsion contribution as
\be
\Gamma^\alpha{}_{\mu\nu}=\Gammab^\alpha{}_{\mu\nu}+K^\alpha{}_{\mu\nu},
\label{eq:Gammadec}
\ee
where
\be
K^\alpha{}_{\mu\nu}=\frac12 \Big(T^\alpha{}_{\mu\nu}+T_{\mu}{}^\alpha{}_\nu+T_\nu{}^\alpha{}_\mu\Big).
\ee
This decomposition can be plugged into the Lagrangian \eqref{eq:PGaction} so we can unveil the presence of pathological terms in a background independent approach just by looking at the interactions of the different torsion components. In order to avoid ghosts already for the graviton when the torsion is turned off, we will impose the recovery of the Gauss-Bonnet term in the limit of vanishing torsion. In four dimensions this amounts to using the topological nature of the Gauss-Bonnet term to remove one of the parameters. More explicitly, we have
\begin{eqnarray}
\mathcal{L}_{\rm PG}\big\vert_{T=0}=\frac12\mpl^2\bar{R}+\left(b_{2}+b_{3}+\frac{b_{4}}{2}\right)\bar{R}_{\mu\nu\rho\sigma}\bar{R}^{\mu\nu\rho\sigma}+\left(b_{5}+b_{6}\right)\bar{R}_{\mu\nu}\bar{R}^{\mu\nu}+b_{1}\bar{R}^{2},
\label{eq:PGactionzeroT}
\end{eqnarray}
so the Gauss-Bonnet term for the quadratic sector is recovered upon requiring
\begin{equation}
\label{con1}
b_{5}=-4b_{1}-b_{6},\;\;\;b_{4}=2(b_{1}-b_{2}-b_{3}),
\end{equation}
that we will assume from now on unless otherwise stated. The parameter $b_1$ will play the role of the coupling constant for this Gauss-Bonnet term. In $d=4$ dimensions this parameter is irrelevant, but it is important for $d>4$.

\subsection{Ghosts in the vector sector}

We will start by looking at the vector sector containing the trace $T_\mu$ and the axial component $S_\mu$ of the torsion, whereas we will neglect the pure tensor piece for now. If we plug \eqref{eq:Gammadec} into the Lagrangian \eqref{eq:PGaction} we obtain
\bea
\Lag_{\rm v}&=&-\frac29\big(\kappa-\beta\big)\mT_{\mu\nu}\mT^{\mu\nu}+\frac{1}{72}\big(\kappa-2\beta\big)\mS_{\mu\nu}\mS^{\mu\nu}+\frac12m_T^2T^2+\frac12m_S^2S^2+\frac{\beta}{81} S^2 T^2\nonumber\\
&&+\frac{4\beta-9b_2}{81}\Big[(S_\mu T^\mu)^2+3S^\mu S^\nu \nablab_\mu T_\nu\Big]+\frac{\beta}{54} S^2\nablab_\mu T^\mu+\frac{\beta-3b_2}{9}S^\mu T^\nu\nablab_\mu S_\nu \nonumber\\
&&+\frac{\beta-3b_2}{12} (\nablab_\mu S^\mu)^2+\frac{\beta}{36}\Big(2\Gb^{\mu\nu}S_\mu S_\nu+\Rb S^2\Big),
\label{eq:PGaction2}
\eea
where $\mT_{\mu\nu}=2\partial_{[\mu} T_{\nu]}$ and $\mS_{\mu\nu}=2\partial_{[\mu} S_{\nu]}$ are the field strengths of the trace and axial vectors respectively and we have defined
\bea
\kappa&=&4b_1+b_6\, ,\\
\beta&=&b_1+b_2-b_3\, ,\\
m_T^2&=&-\frac13\big(2-2a_1+a_2-3a_3\big)\mpl^2\, ,\\
m_S^2&=&\frac{1}{24}\big(1-4a_1-4a_2\big)\mpl^2.
\eea
To arrive at the final expression in \eqref{eq:PGaction2} we have used the Bianchi identities to eliminate terms containing $\Rb_{\mu\nu\rho\sigma}\epsilon^{\alpha\nu\rho\sigma}$ and express $\Rb_{\mu\nu\rho\sigma}\Rb^{\mu\rho\nu\sigma}=\frac12\Rb_{\mu\nu\rho\sigma}\Rb^{\mu\nu\rho\sigma}$. We have also dropped the Gauss-Bonnet invariant of the Levi-Civita connection and the total derivative $\varepsilon_{\mu\nu\alpha\beta} \mS^{\mu\nu}\mT^{\alpha\beta}$. Furthermore, we have performed a few integrations by parts and used the commutator of covariant derivatives. Let us notice that the parameter $b_1$ does not play any role and can be freely fixed because it simply corresponds to the irrelevant Gauss-Bonnet coupling constant.

The Lagrangian \eqref{eq:PGaction2} features some interesting properties. Firstly, if we look at the pure trace sector $T_\mu$, we see that it does not exhibit non-minimal couplings. This is an accidental property in four and lower dimensions, while in higher dimensions the trace vector does couple to the curvature. To show this property more explicitly, the Lagrangian for the pure trace sector in an arbitrary dimension $d$ is given by
\bea
\Lag^{d}_{T}&=&-\frac{d-2}{(d-1)^2}\left(\frac{d-2}{2}\kappa-\beta\right)\mT_{\mu\nu}\mT^{\mu\nu}+\frac12m_T^2(d)T^2\nonumber\\
&&+b_1\frac{(d-4)(d-3)(d-2)}{(d-1)^3}\left[\Big(T^4-4T^2\nablab_\mu T^\mu\Big)+4\frac{d-1}{d-2}\Gb_{\mu\nu} T^\mu T^\nu\right]
\label{eq:Td}
\eea
with
\be
m_T^2(d)=\frac{1}{1-d}\Big[(d-2)-2a_1+a_2+(1-d)a_3\Big]\mpl^2.
\ee
As advertised, all the interactions trivialise\footnote{Notice that $b_1$ controls the Gauss-Bonnet term also for arbitrary dimension $d$, so the trace interactions only contribute if the Gauss-Bonnet is also present, which is dynamical for $d>4$.} in $d=4$ dimensions. It is remarkable however that the non-gauge-invariant derivative interaction $T^2\nablab_\mu T^\mu$ is of the vector-Galileon type and the non-minimal coupling is only to the Einstein tensor, which is precisely one of the very few ghost-free couplings to the curvature for a vector field (see e.g. \cite{Jimenez:2016isa}).  The obtained result agrees with the findings in \cite{Jimenez:2015fva,Jimenez:2016opp} where a general connection determined by a vector field that generates both torsion and non-metricity was considered.  The decoupled trace sector of the PGTs coincides with the connection considered there in the relevant quadratic curvature terms for the particular case without non-metricity. 

Let us now return to the full vector sector. Unlike the torsion trace, the axial component $S_\mu$ exhibits very worrisome terms that appear in three forms:
\begin{itemize}
\item  The perhaps most evidently pathological term is $(\nablab_\mu S^\mu)^2$ that introduces a ghostly dof associated to the temporal component\footnote{Let us remark that, although it looks like a usual gauge-fixing term, it is fully physical in the present case where there is no $U(1)$ gauge symmetry. Consequently, this term makes the temporal component propagate and gives rise to a ghost that cannot be removed from the spectrum by restricting the Hilbert space on the grounds of a gauge symmetry.} $S_0$, so we need to get rid of it by imposing $\beta=3b_2$. This constraint has already been found in the literature in order to guarantee a stable spectrum on Minkowski.

\item {The non-minimal couplings to the curvature, though less obviously than $(\nablab_\mu S^\mu)^2$, are also known to lead to ghostly dof's \cite{Himmetoglu:2008zp,Jimenez:2008sq,ArmendarizPicon:2009ai,Himmetoglu:2009qi}. The presence of these instabilities are apparent in the metric field equations where again the temporal component will enter with second derivatives, thus revealing its problematic dynamical nature. As mentioned above, an exception is the coupling to the Einstein tensor that avoids generating second derivatives of the temporal component thanks to its divergenceless property. For this reason we have explicitly separated the non-minimal coupling to the Einstein tensor in \eqref{eq:PGaction2}. It is therefore clear that we need to impose the additional constraint $\beta=0$ to guarantee the absence ghosts, which, in combination with the above condition $\beta=3b_2$, results in $\beta=b_2=0$. }

\item In addition to the two previous worrisome terms, there are other interactions with a generically pathological character schematically given by $S^2\nabla T$ and $ST\nabla S$. Although these may look like safe vector Galileon-like interactions, the fact that they contain both sectors actually makes them dangerous. This can be more easily understood by introducing St\"uckelberg fields and taking an appropriate decoupling limit, so we effectively have $T_\mu\rightarrow\partial_\mu T$ and $S_\mu\rightarrow\partial_\mu S$ with $T$ and $S$ the scalar and pseudo-scalar St\"uckelbergs. The interactions in this limit become of the form $(\partial S)^2\partial^2 T$ and $\partial T\partial S\partial^2T$ that, unlike the pure Galileon interactions, generically give rise to higher order equations of motion and, therefore, Ostrogradski instabilities\footnote{It is not difficult to check that an interaction for two scalars $\phi$ and $\psi$ of the form $\mathcal{K}^{\mu\nu}(\partial\phi,\partial\psi)\partial_\mu\partial_\nu\phi$ only avoids Ostrogradski instabilities if $\mathcal{K}^{00}$ does not contain time-derivatives of the scalars.}. We will show the problematic nature of these interactions in the explicit example given in Sec. \ref{Sec:Example}.

\end{itemize}

The additional constraint $\beta=0$ conforms the crucial obstruction for stable PGTs. Let us notice first that this new constraint genuinely originates from the quadratic curvature interactions in the PGT Lagrangian that induces the non-minimal couplings between the axial sector and the gravitons, as well as the problematic non-gauge-invariant derivative interactions. Moreover, it cannot be obtained  from a perturbative analysis on a Minkowski background because, in that case, these interactions will only enter at cubic and higher orders so that the linear analysis is completely oblivious to it.

We can see that the two stability conditions not only remove the obvious pathological interactions aforementioned, but they actually eliminate all the interactions and only leave the free quadratic part
\bea
\Lag_{\rm v}\big\vert_{b_2,\beta=0}&=&-\frac{2}{9}\kappa\mT_{\mu\nu}\mT^{\mu\nu}+\frac12m^2_T T^2+\frac{1}{72}\kappa\mS_{\mu\nu}\mS^{\mu\nu}+\frac12m^2_S S^2
\label{eq:PGaction4}
\eea
where we see that crucially the kinetic terms for $T_\mu$ and $S_\mu$  have the same normalisation but with opposite signs, thus signalling the unavoidable presence of a ghost. We are then led to the only stable possibility of exactly cancelling both kinetic terms and, consequently, the entire vector sector becomes non-dynamical.

%

\subsection{Including the tensor sector}

After showing that the vector sector must trivialise in stable PGTs, we can return to the full torsion scenario by including the pure tensor sector $q^\alpha{}_{\mu\nu}$. Instead of using the general decomposition \eqref{eq:torsiondec}, it is more convenient to work with the torsion directly for our purpose here. We can perform the post-Riemannian decomposition for the theories with a stable vector sector to obtain
\begin{eqnarray}
\mathcal{L}_{\rm stable}&=&\frac12\mpl^2\bar{R}+b_{1}\GB+\frac12\mpl^2\Big(a_{1}T_{\mu\nu\rho}T^{\mu\nu\rho}+a_{2}T_{\mu\nu\rho}T^{\nu\rho\mu}+a_{3}T_\mu T^\mu\Big).
\end{eqnarray}
The first term is just the usual Einstein-Hilbert Lagrangian, while the second term corresponds to the topological Gauss-Bonnet invariant for a connection with torsion, so we can safely drop it in four dimensions and, consequently, the first two terms in the above expression simply describe GR. The rest of the expression clearly shows the non-dynamical nature of the full torsion so that having a stable vector sector also eliminates the dynamics for the tensor component, thus making the full connection be an auxiliary field. We can straightforwardly integrate the connection out and, similarly to the Einstein-Cartan theories, the resulting effect will be the generation of interactions for fermions that couple to the axial part of the connection. From an EFT perspective, the effect will simply be a shift in the corresponding parameters of those interactions with no observable physical effect whatsoever.

\subsection{Explicit example}\label{Sec:Example}
The exact post-Riemannian expansion of the quadratic PGTs has unveiled the generic presence of ghosts and how their avoidance results in the trivialisation of the whole torsion sector. We will now show how the ghosts appear and we will rederive the same conclusions by working out an explicit example. This is important to guarantee the absence of hidden constraints that could secretly render the theory stable even if the Lagrangian contains dangerous-looking operators. In this respect, it is important to bear in mind that worrisome terms can be generated from perfectly healthy interactions via field redefinitions (see e.g. the related discussion in \cite{Jimenez:2019hpl}) so we must make sure that the terms arising in the quadratic PGTs do not correspond to some obscure formulation of well-behaved theories. There is no obvious reason to expect any such mechanism at work for PGTs and in fact we will demonstrate that this is not the case in a very simple setup.

In order to prove the dynamical nature of  $S_0$ we will consider a homogeneous vector sector on a cosmological background described by the FLRW metric\footnote{Actually, it would be sufficient to work on a Minkowski background. We prefer however to use a general cosmological background to not trivialise any interaction in \eqref{eq:PGaction2} and to explicitly show the irrelevant role of the curvature for our analysis.}
\be
\dd s^2=a^2(t)\big(-\dd t^2+\dd\vec{x}^2\big).
\label{eq:FLRW}
\ee
The tensor sector is kept trivial so we only care about the vector components. It is straightforward to see from \eqref{eq:PGaction2} that $T_0$ is always an auxiliary field. Its equation of motion for the considered configuration is given by
\bea
0&=&\Big[-27m_T^2a^2+2(\beta-3b_2)S_0+\frac23\beta S_z^2\Big]T_0+\frac23(9b_2-4\beta)S_0\vec{S}\cdot\vec{T}+6(3b_2-\beta)HS_0^2\nonumber\\
&&-2\beta H\vec{S}^2+\frac{3}{2}(3b_2-2\beta)(S_0^2)'+\frac\beta2(\vec{S}^2)',
\eea
so we can solve for it and integrate it out from the action. After performing a few integrations by parts and choosing $\vec{T}$ and $\vec{S}$ aligned with the $z-$axis, we can compute the corresponding Hessian from the resulting Lagrangian, whose cumbersome form is not very illuminating so we omit it here. The expression for the Hessian is rather simple and reads
\bea
\mathcal{H}_{ij}&=&\frac{\delta \mS_{\rm B}}{\delta \dot{X}_i\delta\dot{X}_j}=
\left(
\begin{array}{ccc}
  \lambda_1& \tilde{\lambda} &  0 \\
  \tilde{\lambda}& \lambda_2  &   0\\
 0 &0   &   \frac89(\kappa-\beta)
\end{array}
\right)\nonumber
\eea
with $\vec{X}=(S_0,T_z,S_z)$ and we have defined
\bea
\lambda_1&=&\frac{\beta-3b_2}{6}+\frac{(3b_2-2\beta)^2 S_0^2}{81 m_T^2 a^2 + 6 (3 b_2 - \beta) S_0^2 - 2 \beta S_z^2},\nonumber\\
\lambda_2&=&\frac{1}{18}\left(\beta-\kappa+\frac{81 m_T^2 a^2 + 6 (3 b_2 - \beta) S_0^2 }{81 m_T^2 a^2 + 6 (3 b_2 - \beta) S_0^2 - 2 \beta S_z^2}\right),\nonumber\\
\tilde{\lambda}&=&\frac13 \frac{(3b_2-2\beta)\beta }{81 m_T^2 a^2 + 6 (3 b_2 - \beta) S_0^2 - 2 \beta S_z^2}S_0S_z\,.
\eea
The presence of constraints can be determined by computing the determinant of the Hessian. It is easy to see that, in general, $\det \mathcal{H}_{ij}\neq0$, thus guaranteeing the absence of any additional constraints so that $S_0$ indeed represents a fully propagating dof. In order to ensure the presence of constraints we need to solve the equation $\det \mathcal{H}_{ij}=0$ for arbitrary values of the fields. By solving this equation  we recover the conditions $\beta=b_2=0$ and the Hessian reduces to 
\bea
\mathcal{H}_{ij}=
\left(
\begin{array}{ccc}
  0&0 &  0 \\
  0& -\frac{1}{18}\kappa &   0\\
 0 &0   &   \frac89\kappa
\end{array}
\right)\nonumber
\eea
that is trivially degenerate and ensures a non-propagating $S_0$. Moreover, we also see the ghostly nature of either $T_\mu$ or $S_\mu$ since the non-vanishing eigenvalues have opposite signs. These results confirm the conclusions reached from the exact post-Riemannian analysis.

\section{Stabilising quadratic PGT's}
The precedent section has been devoted to showing the presence of ghosts in general quadratic PGTs. Although this is a drawback for generic theories, we will now show how to avoid the presence of the discussed instabilities by following different routes. In particular, we will show specific class of ghost-free theories and how to stabilise the vector sector by adding suitable operators of the same dimensionality as those already present in the quadratic PGTs. Before showing the ghost-free theories, let us discuss the failure of some approaches that may seem promising at first sight.

\subsection{Dead routes}

\subsubsection{Trivial torsion-free limit}
An important requirement in order to recover a known ghost-free graviton sector at zero torsion was to impose the recovery of the Gauss-Bonnet term in that regime\footnote{While in $d=4$ the Gauss-Bonnet term is a total derivative, we have seen its non-trivial role in higher dimensions in \eqref{eq:Td}.}. It is natural to wonder if the condition of obtaining Gauss-Bonnet in the vanishing torsion limit is too restrictive and it could be relaxed. Thus, we will explore now what happens if we impose a different ghost-free vanishing torsion limit. Perhaps an obvious approach would be to allow only for objects that identically vanish for a Levi-Civita connection such as the antisymmetric Ricci tensor $R_{[\mu\nu]}$. One can however verify that the opposite parity of $T_\mu$ and $S_\mu$ precisely prevents ghost-freedom. More explicitly, we have that $R_{[\mu\nu]}=\frac13\mT_{\mu\nu}+\frac{1}{12}\varepsilon_{\mu\nu\alpha\beta}\mS^{\alpha\beta}$ so 
\be
R_{[\mu\nu]}R^{[\mu\nu]}=\frac19\left(\mT_{\mu\nu} \mT^{\mu\nu}-\frac{1}{16}\mS_{\mu\nu} \mS^{\mu\nu}+\varepsilon_{\mu\nu\alpha\beta}\mT^{\mu\nu}\mS^{\alpha\beta}\right).
\ee
The last term is topological and can be safely dropped since it will not contribute to the equations of motion. We thus see that the quadratic PGT only containing the antisymmetric Ricci tensor necessarily produces a ghost because the kinetic terms of the trace and axial vectors enter with opposite signs\footnote{This result is contradiction with the findings in \cite{Vasilev:2017twr}. The disagreement is due to a missing $-1$ factor in \cite{Vasilev:2017twr}. Once this factor is corrected the results are in perfect agreement. We thank the authors of  \cite{Vasilev:2017twr} for their help in clarifying the disagreement.}. This can also be checked by making $b_1=b_2=b_3=b_4=0$ and $b_5=-b_6$ in \eqref{eq:PGaction2}.

\subsubsection{Parity-violating terms}
It has been shown in the literature that including parity-violating terms may help with the stability properties of these theories \cite{Blagojevic:2018dpz}. Hence, it is natural to study the effects of such terms in relation with our no-go result. Since only curvature parity violating terms can contribute kinetic terms for the vector sector, we will restrict to them here. In particular, the following parity-violating terms can be added to the Lagrangian
\begin{eqnarray}
\mathcal{L}_{\rm odd}=&&\varepsilon^{\mu\nu\rho\sigma}\left(d_{1}RR_{\mu\nu\rho\sigma}+d_{2}R_{\alpha\beta\mu\nu}R^{\alpha\beta}{}_{\rho\sigma}+d_{3}R_{\mu\nu\alpha\beta}R^{\alpha\beta}{}_{\rho\sigma}\right).
\end{eqnarray}
We have not included $\varepsilon^{\mu\nu\rho\sigma}R_{\mu\nu\alpha\beta}R_{\rho\sigma}{}^{\alpha\beta}$ because it corresponds to the Pontryagin topological invariant. Moreover, adding the above terms  still respects the required Gauss-Bonnet limit for vanishing torsion because they either trivialise or  reduce to the Pontryagin invariant for the Levi-Civita connection. By performing a post-Riemannian expansion for these parity breaking terms we obtain, up to integrations by parts, the following contributions for the vector sector:
\begin{eqnarray}
\label{parvector}
\mathcal{L}_{\rm odd,v}&=&\frac{2}{9}\kappa_\star\mS_{\mu\nu}\mT^{\mu\nu}-\frac{2}{9}\gamma\bar{R}S_{\mu}T^{\mu}-\frac{1}{108}\gamma S^2S_{\mu}T^{\mu}+\frac{4}{27}\gamma T^2S_{\mu}T^{\mu}+\frac{1}{3}\gamma\bar{R}\bar{\nabla}_{\mu}S^{\mu}
\nonumber
\\
&+&\frac{1}{72}\gamma S^{2}\bar{\nabla}_{\mu}S^{\mu}-\frac{2}{9}\gamma T^2\bar{\nabla}_{\mu}S^{\mu}-\frac{4}{9}\gamma S_{\mu}T^{\mu}\bar{\nabla}_{\nu}T^{\nu}+\frac{2}{3}\gamma\bar{\nabla}_{\mu}S^{\mu}\bar{\nabla}_{\nu}T^{\nu},
\end{eqnarray}
where $\kappa_\star=2d_2-d_3$ and $\gamma=3d_{1}+2d_{2}+d_{3}$. We can see that all the potentially dangerous terms involving non-gauge invariant derivatives of the vectors and non-minimal couplings can be eliminated by setting $\gamma=0$. For parameters satisfying this condition, the parity breaking Lagrangian for the vector sector reduces to
\begin{equation}
\Lag_{\rm odd,v}\big\vert_{\gamma=0}=\frac{2}{9}\kappa_{\star}\mS_{\mu\nu}\mT^{\mu\nu}.
\label{eq:Lparmix}
\end{equation}
This mixing between the field strengths of the trace and axial components cannot stabilise the ghost-like instability of the parity preserving theory for any choice of $\kappa_\star$. The kinetic matrix adding the parity violating term to \eqref{eq:PGaction4} is
\[
\hat{K}=\frac{1}{9}\left(
\begin{array}{cc}
  -2\kappa&   \kappa_\star   \\
\kappa_\star  &   \kappa/8   
\end{array}
\right).
\]
If we compute the determinant we obtain 
\be
\det\hat{K}=-\frac{\kappa^2+4\kappa_\star^2}{324}
\ee
 which is negative for any choice of the parameters\footnote{This is just an example of how adding non-diagonal terms in the kinetic matrix cannot turn a ghost into a healthy mode. Rather, the opposite, very large off-diagonal contributions could turn a healthy field into a ghost.}. This clearly signals that the two eigenvalues have opposite signs and, as a consequence, the ghost will always be present so the addition of parity breaking terms does not help rendering the theory ghost-free. This could have been anticipated because dimension 4 and parity violating operators can only generate a term like \eqref{eq:Lparmix} which, as we have shown, cannot fix the ghostly nature of the parity-preserving sector.

\subsubsection{Alternative ghost-free vanishing torsion regime}
Among all the quadratic gravity theories in the metric formalism, which generically contain ghosts, it is well-known that the particular case of a correction $\Rb^2$ to the Einstein-Hilbert action is a safe quadratic modification that introduces an extra healthy scalar. Let us then explore the theories that reduce to this ghost-free action at vanishing torsion so we now impose 
\be
b_5=-b_6,\quad\text{and}\quad b_4=-2(b_2+b_3),
\ee
that only leaves $b_1\Rb^2$ in \eqref{eq:PGactionzeroT}. One can quickly be convinced that the $\Rb^2-$limit  does not help much for the ghost-freedom requirement in the vector sector. In fact, rather the opposite, i.e., it leads to an even more pathological Lagrangian than the theory with the Gauss-Bonnet limit because now also the trace sector $T_\mu$ features ghostly interactions. In particular, we find
\be
\Lag\supset4b_1\left[\frac13\Rb T^2+R\nabla_\mu T^\mu+(\nabla_\mu T^\mu)^2\right]
\ee
which requires $b_1=0$ to avoid ghosts. However, this precisely corresponds to removing the $\Rb^2-$ term so that we in turn recover the theory with a GR limit analysed above. There is nevertheless a special choice of parameters  that avoids this negative result and that we explore in detail in the next section.

\subsection{$R^2$ theories}
The specific parameters choice that leads to a stable class of theories corresponds to further restricting the quadratic curvature sector to be exactly the Ricci scalar square of the full connection $R^2$. This theory will evidently have the $\Rb^{2}-$limit at vanishing torsion, but it avoids the ghostly interactions that originate from the other Riemann contractions as we show in the following. We thus set the parameters to $b_2=b_3=b_4=b_5=b_6=0$ and $b_1\neq0$ so we will consider the particular PG Lagrangian
\begin{align}
\mathcal{L}=&\frac12\mpl^2\Big({R}+a_{1}T_{\mu\nu\rho}T^{\mu\nu\rho}
+a_{2}T_{\mu\nu\rho}T^{\nu\rho\mu}+a_{3}T_\mu T^\mu\Big)+b_{1}{R}^{2}.
\label{eq:PGactionR2}
\end{align}
This specific Lagrangian and its non-pathological character was already found in \cite{Hecht:1996np,Yo:1999ex} by analysing its well-posedness and Hamiltonian structure. Our  approach here will confirm these results by a different procedure and will give further insights. The idea is to rewrite the Lagrangian in a form where the additional scalar is made explicit. As usual, we start by performing a Legendre transformation in order to recast the Lagrangian in the more convenient form
\be
\Lag=\frac12\mpl^2\varphi+b_1\varphi^2+\chi(R-\varphi) +\frac12m_T^2 T^2+\frac12m_S^2S^2,
\ee
where we have introduced the non-dynamical fields $\chi$ and $\varphi$ and we have left the pure tensor sector $q^\alpha{}_{\mu\nu}$ out for the moment, although we will come back to its relevance later. Upon use of the field equation for $\chi$ we recover the original Lagrangian, while the equation for $\varphi$ yields
\be
\varphi=\frac{2\chi-\mpl^2}{4b_1}
\ee
that gives $\varphi$ as a function of $\chi$. We can now use the post-Riemannian expansion of the Ricci scalar
\be
R=\Rb+\frac{1}{24}S^2-\frac{2}{3}T^2+2\nablab_\mu T^\mu
\label{eq:Ricciscalar}
\ee
to express the Lagrangian in the following suitable form
\bea
\Lag&=&\chi\left(\Rb+\frac{1}{24}S^2-\frac{2}{3}T^2+2\nablab_\mu T^\mu\right)-\frac{\big(2\chi-\mpl^2\big)^2}{16b_1}+\frac12m_T^2T^2+\frac12 m_S^2S^2.
\label{eq:PGactionR22}
\eea
The equation for the axial part imposes $S_\mu=0$, while the trace part yields
\be
T_\mu=\frac{2\partial_\mu\chi}{m_T^2-\frac43 \chi}
\ee
which shows that $T_\mu$ can only propagate a scalar\footnote{An analogous result was obtained in \cite{Ozkan:2015iva} by considering $f(R)$ theories where the Ricci scalar is replaced by $R\rightarrow R+A^2+\beta \nablab_\mu A^\mu$ with $A_\mu$ a vector field and in \cite{Jimenez:2015fva} within the context of geometries with vector distortion. Interestingly, these scenarios provide a realisation of the $\alpha$-attractor model of inflation.} since it can be expressed as $T_\mu=\partial_\mu\tilde{\chi}$ with
\be
\tilde{\chi}=-\frac32\log\Big\vert3m_T^2-4\chi\Big\vert.
\ee
The theory is then equivalently described by the action\footnote{We prefer to give the action to make explicit the conformal factors coming from the volume element.}
\be
\mS=\int\dd^4x\sqrt{-g}\left[\chi\Rb-\frac{2(\partial\chi)^2}{m_T^2-\frac{4}{3}\chi}-\frac{\big(2\chi-\mpl^2\big)^2}{16b_1}\right]
\label{eq:BDPGT}
\ee
which reduces to a simple scalar-tensor theory of a generalised Brans-Dicke type with a field dependent Brans-Dicke parameter:
\be
\omega_{\rm BD}(\chi)=\frac{2\chi}{m_T^2-\frac43\chi}\,.
\ee
 This result actually extends to arbitrary $f(R)$ extensions of PGTs, the only difference with respect to \eqref{eq:BDPGT} being the specific form of the potential for $\chi$. A noteworthy feature of the resulting Lagrangian is the singular character of the massless limit $m_T^2=0$ that gives $\omega_{\rm BD}(m_T^2=0)=-3/2$, precisely the value that makes the scalar field non-dynamical. This is also the case for the Palatini formulation of $f(R)$ theories where the scalar is non-dynamical (see e.g. \cite{Olmo:2011uz} and references therein). For any other value of the mass, the scalar field is fully dynamical. We can see this more explicitly by performing the conformal transformation $\gt_{\mu\nu}=\frac{2\chi}{\mpl^2}g_{\mu\nu}$ that brings the action into the Einstein frame 
 \begin{align}
\mS=\int\dd^4x\sqrt{-\gt}&\left[\frac12\mpl^2\Rt-\frac{3m_T^2\mpl^2}{4\chi^2(m_T^2-\frac43\chi)}(\partial\chi)^2-\frac{\mpl^2}{8  b_1}\left(1-\frac{\mpl^2}{2\chi}\right)^2\right].
\label{eq:R2Einstein}
 \end{align}
In this frame it becomes apparent that the scalar $\chi$ loses its kinetic term for $m_T^2=0$. Of course, this feature can be related to the breaking of a certain conformal symmetry by the mass term. If we perform a conformal transformation of the metric together with a projective transformation of the torsion\footnote{The torsion transformation is $T^\alpha{}_{\mu\nu}\rightarrow T^\alpha{}_{\mu\nu}-2\delta^\alpha_{[\mu}\partial_{\nu]}\Omega$ that gives the transformation for the vector trace quoted in the main text, while the axial and pure tensor pieces remain invariant. See e.g. \cite{Obukhov:1982zn,HelayelNeto:1999tm,Shapiro:2001rz} for interesting discussions on conformal transformations involving torsion.} given by
\be
 g_{\mu\nu}\rightarrow e^{2\Omega}g_{\mu\nu},\quad  T_\mu\rightarrow T_\mu+3\partial_\mu\Omega
 \ee
with $\Omega$ and arbitrary function, we have that the Ricci scalar transforms as $R\rightarrow e^{-2\Omega}R$. Thus, we have that the Lagrangian  \eqref{eq:PGactionR22} is invariant under the above transformations supplemented with $\chi\rightarrow e^{-2\Omega}\chi$ except for the mass term\footnote{Actually, the potential for $\chi$ also breaks the conformal invariance, but since it does not affect the dynamical nature of $\chi$ we can neglect it for our discussion here.}. Thus, for $m_T^2=0$, the fact that the torsion is given in terms of the gradient of $\chi$  together with the discussed symmetry allows to completely remove the kinetic terms for $\chi$ by means of a conformal transformation. The mass however breaks this symmetry and, consequently, we recover the dynamical scalar described by \eqref{eq:R2Einstein}. Furthermore, the mass $m_T^2$ also determines the region of ghost freedom for the theory. If $m_T^2>0$ we have an upper bound for the scalar field that must satisfy $\chi<\frac34m_T^2$ in order to avoid the region where it becomes a ghost. On the other hand, if $m_T^2<0$, the scalar field is confined to the region $\chi>\frac34m_T^2$. For the potential to be bounded from below we only need to have $b_1>0$. These conditions have been summarised in Table \ref{table}.
 
It may be worth noticing that the absence of ghosts in the $R^2$-theories is due to the removal of the Maxwell kinetic terms for the vector sector. By inspection of the Ricci scalar \eqref{eq:Ricciscalar} we see that only the trace $T_\mu$ enters with derivatives and only through the divergence $\nablab_\mu T^\mu$. As it is well-known this is precisely the dual of the usual Maxwell-like kinetic term for the dual 3-form field so the theory can be associated to a massive 3-form which propagates one dof\footnote{See e.g. \cite{Germani:2009iq, Koivisto:2009sd,Koivisto:2009ew} for some cosmological applications of 3-forms.}. This dof can be  identified with the scalar that we have found. Just like the $U(1)$ gauge symmetry of the Maxwell terms is crucial for the stability of vector theories, the derivative term $\nablab_\mu T^\mu$ has the symmetry $T^\mu\rightarrow T^\mu+\epsilon^{\mu\nu\rho\sigma}\partial_{\nu} \theta_{\rho\sigma}$ for an arbitrary $\theta_{\rho\sigma}$ that plays a crucial role for guaranteeing the stability of the theories. Of course, this symmetry is inherited from the gauge symmetry of the dual 3-form.

Let us finally notice that including the tensor sector $q^\alpha{}_{\mu\nu}$ does not change the final result because one can check that, similarly to the axial part, it only enters as an auxiliary field whose equation of motion imposes $q^\alpha{}_{\mu\nu}=0$. To see this more clearly, we can give the full post-Riemannian expansion of the Ricci scalar including the tensor piece
\be
R=\Rb+\frac{1}{24}S^2-\frac{2}{3}T^2+2\nablab_\mu T^\mu+\frac12q_{\mu\nu\rho}q^{\mu\nu\rho},
\ee
so it is clear that its contribution to the Lagrangian \eqref{eq:PGactionR22} gives rise to the equation of motion $\chi q_{\mu\nu\rho}=0$ which, for $\chi\neq0$,  trivialises the tensor component. The same will apply to theories described by an arbitrary function $f(R)$ so one can safely neglect the tensor sector for those theories as well.


\subsection{Holst square theories}
 In the precedent section we have seen how to obtain a non-trivial quadratic PGT that propagates an extra-scalar and this can be ultimately related to the absence of Maxwell-like terms for the vector sector. We can then ask whether there is some non-trivial healthy theory described by \eqref{eq:PGaction2} where the scalar is associated to the axial vector rather than to the trace. The answer is affirmative and in order to obtain it we simply need to impose the vanishing of the Maxwell kinetic terms that results in the conditions:
\be
\kappa=0\quad {\text{and}}\quad \beta=0.
\ee
Under these conditions, after performing a few integrations by parts and dropping a Gauss-Bonnet term, the Lagrangian reads
\bea
\Lag_{\rm Holst}&=&\frac12\mpl^2\Rb+\frac12m_T^2 T^2+\frac12m_S^2S^2
\nonumber
\\
&&+\alpha\left[(\nablab_\mu S^\mu)^2-\frac43 S_\mu T^\mu\nablab_\nu S^\nu+\frac49(S_\mu T^\mu)^2\right] 
\label{eq:Holst1}
\eea
with $\alpha\equiv-\frac{b_2}{4}$. It is apparent that we obtain the same structure as in the $R^2$ case but now for the axial part. This is not an accident and it can be understood from the relation of the resulting Lagrangian with the Holst term\footnote{Although this term is commonly known as the Holst term, due to the research article of Soren Holst in 1995 \cite{Holst:1995pc}, in the context of torsion gravity it was first introduced by R. Hojman {\emph{et. al.}} in 1980 \cite{Hojman:1980kv}.} \cite{Hojman:1980kv,Holst:1995pc} that is given by $\mathcal{H}\equiv\epsilon^{\mu\nu\rho\sigma} R_{\mu\nu\rho\sigma}$ and whose post-Riemannian expansion is
\be
\mathcal{H}=\frac23 S_\mu T^\mu-\nablab_\mu S^\mu
\label{eq:HoldstPR}
\ee
where we have used that $\epsilon^{\mu\nu\rho\sigma} \Rb_{\mu\nu\rho\sigma}=0$ by virtue of the Bianchi identities. It is then obvious that the Lagrangian can be written as
\be
\Lag_{\rm Holst}=\frac12\mpl^2\Rb+\frac12m_T^2 T^2+\frac12m_S^2S^2
+\alpha \mathcal{H}^2 .
\label{eq:Holst2}
\ee
This particular PGT was identified in \cite{Hecht:1996np} as an example of a theory with dynamical torsion described by a scalar with a well-posed initial value problem. We will unveil the nature of this scalar by proceeding in an analogous manner to the $R^2$ theories. For that we first introduce an auxiliary field $\phi$ to rewrite \eqref{eq:Holst2} as
\bea
\Lag_{\rm Holst}&=&\frac12\mpl^2\Rb+\frac12m_T^2 T^2+\frac12m_S^2S^2-\alpha\phi^2+2\alpha\phi\epsilon^{\mu\nu\rho\sigma} R_{\mu\nu\rho\sigma}.
\label{eq:Holst3}
\eea
We see that the resulting equivalent Lagrangian corresponds to the addition of a Holst term where the Barbero-Immirzi parameter has been promoted to be a pseudo-scalar field. As we will show now, this pseudo-scalar is dynamical and corresponds to the $0^-$ mode identified in \cite{Hecht:1996np}. The massless theory with $m_ T^2=m_S^2=0$ and without the $\phi^2$ potential has been considered as extensions of GR inspired by Loop Quantum Gravity \cite{Taveras:2008yf,Calcagni:2009xz}. We can now introduce the post-Riemannian expansion \eqref{eq:HoldstPR} into the Lagrangian, so we have
\bea
\Lag_{\rm Holst}&=&\frac12\mpl^2\Rb+\frac12m_T^2 T^2+\frac12m_S^2S^2-\alpha\phi^2+2\alpha\phi\left(\frac23 S_\mu T^\mu-\nablab_\mu S^\mu\right).
\label{eq:Holst4}
\eea
The equations for $S^\mu$ and $T^\mu$ are 
\bea
m_S^2S_\mu+\frac{4\alpha\phi}{3}T_\mu+2\alpha\partial_\mu\phi=0,\label{eq:S}\\
m_T^2T_\mu+\frac{4\alpha\phi}{3}S_\mu=0,
\eea
respectively. For $m_T^2\neq0$\footnote{The singular value $m_T^2=0$ leads to uninteresting theories where all the dynamics is lost so we will not consider it any further here. The same conclusion was reached in \cite{Hecht:1996np}.} we can algebraically solve these equations as
\bea
T_\mu&=&-\frac{4\alpha\phi}{3m_T^2}S_\mu,\\
S_\mu&=&-\frac{2\alpha\partial_\mu\phi}{m_S^2-\left(\frac{4\alpha\phi}{3m_T}\right)^2},\label{eq:solS}
\eea
that we can plug into the Lagrangian to finally obtain
\be
\Lag_{\rm Holst}=\frac12\mpl^2\Rb-\frac{2\alpha^2}{m_S^2-\left(\frac{4\alpha\phi}{3m_T}\right)^2}(\partial\phi)^2-\alpha\phi^2.
\label{eq:Holst4}
\ee
This equivalent formulation of the theory with all the auxiliary fields integrated out manifestly exposes the presence of a propagating pseudo-scalar field. The parity invariance of the original Lagrangian translates into a $\mathbb{Z}_2$ symmetry in the pseudo-scalar sector. The obtained result is also valid for theories described by an arbitrary function of the Holst term and considering different functions leads to different potentials for the pseudo-scalar $\phi$. Furthermore, although we have only considered the vector sector of the torsion, including the pure tensor part $q^\alpha{}_{\mu\nu}$ into the picture does not change the conclusions because the latter contributes as
\be
\mathcal{H}=\frac23 S_\mu T^\mu-\nablab_\mu S^\mu+\frac12\epsilon_{\alpha\beta\mu\nu}q_\lambda{}^{\alpha\beta}q^{\lambda\mu\nu}.
\ee
This shows that $q_{\alpha\mu\nu}$ only enters as an auxiliary field whose equation of motion trivialises it, very much as it occurs for the $R^2$ theories.

Let us also point out how the appearance of a (pseudo-)scalar could have been expected by recalling the relation of the Holst term with the Nieh-Yan topological invariant given by
\be
\mathcal{N}\equiv\epsilon^{\mu\nu\rho\sigma}\Big(R_{\mu\nu\rho\sigma}-\frac12 T^\alpha{}_{\mu\nu}T_{\alpha\rho\sigma}\Big).
\ee
In a Riemann-Cartan spacetime it is easy to show that this term is nothing but the total derivative $\mathcal{N}=-\nablab_\mu S^\mu$. The remarkable property of this invariant is that it is linear in the curvature so its square must belong to the class of parity preserving quadratic PGTs, even though $\mathcal{N}$ itself breaks parity. Then, as it happens with other invariants like the Gauss-Bonnet one, including a general non-linear dependence on the invariant is expected to give rise to dynamical scalar modes. In standard Riemannian geometries, the inclusion of an arbitrary function of the Gauss-Bonnet invariant results in a highly non-trivial scalar field with Horndeski interactions (see e.g. \cite{Kobayashi:2011nu}).

The stability requirements for the parameters can now be obtained very easily. From \eqref{eq:Holst4} we can readily conclude that $\alpha$ must be positive to avoid having an unbounded potential from below. On the other hand, the condition to prevent $\phi$ from being a ghost depends on the signs of $m_S^2$ and  $m_T^2$,  which are not defined by any stability condition so far. Accordingly, we can distinguish the following different possibilities:
\begin{itemize}
\item $m_S^2>0$: We then need to have $1-\left(\frac{4\alpha\phi}{3m_Tm_S}\right)^2>0$. For $m_T^2<0$ this is always satisfied, while for $m_T^2>0$ there is an upper bound for the value of the field given by $\vert\phi\vert<\vert\frac{3m_Sm_T}{4\alpha}\vert$.

\item $m_S^2<0$: The ghost-freedom condition is now $1-\left(\frac{4\alpha\phi}{3m_Tm_S}\right)^2<0$, which can never be fulfilled if $m_T^2>0$. If $m_T^2<0$ we instead have the lower bound $\vert\phi\vert>\vert\frac{3m_Sm_T}{4\alpha}\vert$.
\end{itemize}

For a better visualisation we have outlined these ghost-free conditions in Table \ref{table}.\\
We can gain a better intuition on the dynamics of the pseudo-scalar by canonically normalising it. For that purpose we introduce a field $\hat{\phi}$ defined by
\be
\hat{\phi}=\frac{2\alpha}{\sqrt{m_S^2}}\int\frac{\dd\phi}{\sqrt{1-\left(\frac{4\alpha\phi}{3m_Tm_S}\right)^2}}.
\label{eq:canfield}
\ee
For $m_S^2>0$ we obtain
\be
\phi(\hat{\phi})=\frac{3m_Tm_S}{4\alpha}\sin\left(\frac{2\hat{\phi}}{3m_T}\right),
\label{eq:phican1}
\ee
in terms of which the Lagrangian for the pseudo-scalar reads
\be
\Lag_{\hat{\phi}}\vert_{m_S^2>0}=-\frac12(\partial\hat{\phi})^2-V(\hat{\phi}),
\ee
with $V(\hat{\phi})=\alpha\phi^2(\hat{\phi})$. The shape of the potential will crucially depend on the sign of $m_T^2$. Thus, if $m_T^2>0$ we have the oscillatory potential
\be
V(\hat{\phi})=\frac{9 m_T^2m_S^2}{16\alpha}\sin^2\left(\frac{2\hat{\phi}}{3 m_T}\right),\quad\quad m_S^2>0, \;m_T^2>0,
\ee
with a discrete symmetry $\hat{\phi}\rightarrow\hat{\phi}+\frac32nm_T\pi$ with $n\in\mathbb{Z}$ arising from the original upper bound of $\phi$. Notice that the field redefinition \eqref{eq:phican1} guarantees the ghost-free condition $\vert\phi\vert\leq\vert\frac{3m_Tm_S}{4\alpha}\vert$. For $m_T^2<0$ the potential takes instead the form
\be
V(\hat{\phi})=\frac{9\vert m_T^2\vert m_S^2}{16\alpha}\sinh^2\left(\frac{2\hat{\phi}}{3\vert m_T\vert} \right),\quad\quad m_S^2>0, \;m_T^2<0.
\ee

On the other hand, for $m_S^2<0$, we need to have $m_T^2<0$ to avoid ghosts and the integral \eqref{eq:canfield} gives
\be
\phi=\pm\frac{3\vert m_T m_S\vert}{4\alpha}\cosh\left(\frac{4\hat{\phi}}{3\vert m_T\vert}\right),
\ee
where we have fixed the integration constant so that the origin of $\hat{\phi}$ corresponds to the lower bound for $\vert\phi\vert$. The Lagrangian for the canonically normalised field is given by
\be
\Lag_{\hat{\phi}}\vert_{m_S^2<0}=-\frac12(\partial\hat{\phi})^2-\frac{9 m_T^2m_S^2}{16\alpha}\cosh^2\left(\frac{2\hat{\phi}}{3 \vert m_T\vert}\right),\quad\quad m_S^2<0,\; m_T^2<0.
\ee
In all cases, it is straightforward to analyse the corresponding solutions by simply looking at the shape of the corresponding potential. In particular, we see that the small field regime gives an approximate quadratic potential so, provided the mass is sufficiently large\footnote{By large we of course mean relative to the Hubble parameter in the late time universe so that the field can undergo multiple oscillations around the minimum in a Hubble time. This typically requires masses around $m\sim10^{-22}$ eV so they actually represent ultra-light particles from a particle physics perspective.}, the coherent oscillations of the pseudo-scalar can give rise to dark matter \cite{Turner:1983he,PhysRevLett.64.1084,Cembranos:2015oya,Hui:2016ltb} as the misalignment mechanism for axions \cite{Marsh:2015xka} or the Fuzzy Dark Matter models \cite{Hu:2000ke}. A similar mechanism was explored in \cite{Cembranos:2008gj} within pure $R^2$ gravity. On the other hand, it is also possible to generate large field inflationary scenarios or dark energy models if the field slowly rolls down the potential at field values sufficiently far from the minimum.

An important qualitative difference with respect to the $R^2$ theories discussed in the precedent section is that here we have obtained the Lagrangian for the pseudo-scalar already in the Einstein frame, while this was only achieved after performing a conformal transformation to disentangle the scalar field from the Einstein-Hilbert term for the $R^2$ theories. Therefore, while the scalar couples directly to matter in the Einstein frame through a conformal metric for the $R^2$ theories, the pseudo-scalar field of the Holst square theories does not. This could be useful for dark matter and/or dark energy models because they could easily evade local gravity constraints. As a matter of fact, it is noteworthy that the obtained effective potential for the pseudo-scalar field allows for both accelerating cosmologies (that could be used for dark energy or inflation) and dark matter dominated universes. The technical naturalness of the models would of course remain an open challenging issue. A cautionary comment is in order here however because Dirac fermions do couple to the axial part of the connection (see e.g. \cite{Hammond:2002rm,Shapiro:2001rz}). An immediate consequence of this coupling is that actually we would expect to have the dual of the hypermomentum $\Delta_\mu=\delta\mS/\delta S^\mu$ entering on the rhs of \eqref{eq:S}. This means that the solutions for $S_\mu$ and $T_\mu$ in \eqref{eq:solS} should include $\Delta_\mu$ so the final Lagrangian \eqref{eq:Holst4} will feature couplings between the pseudo-scalar $\phi$ and Dirac fermions. Since $\Delta_\mu$ in the equations can be simply generated by the replacement $2\alpha\partial_\mu\phi\rightarrow 2\alpha\partial_\mu\phi+\Delta_\mu$ in \eqref{eq:S}, the explicit computation of the interactions including the axial coupling to the fermions can be easily obtained by making the corresponding replacement in \eqref{eq:Holst4} that yields
\be
\Lag_{\rm Holst}=\frac12\mpl^2\Rb-\frac{(2\alpha\partial_\mu\phi+\Delta_\mu)^2}{m_S^2-\left(\frac{4\alpha\phi}{3m_T}\right)^2}-\alpha\phi^2.
\label{eq:Holst5}
\ee
We then obtain the usual four-point fermion interactions given by $\Delta^2$ that are also generated in e.g. Einstein-Cartan gravity plus a derivative coupling of the pseudo-scalar to the axial current $\Delta_\mu$ carried by the fermions. Interestingly, this derivative coupling can yield an effective mass for the fermion\footnote{Let us recall that the axial current for a fermion $\psi$ has the form $\Delta_\mu\propto \bar{\psi}\gamma_5\gamma_\mu\psi$ so the derivative coupling indeed generates an effective mass.} that depends on the evolution of the pseudo-scalar. A detailed analysis of the phenomenology of these interactions is beyond the scope of this communication, but it is worth noting the possibility that offers this scenario for a natural framework to have dark energy and/or dark matter interacting with neutrinos that could result in some interesting phenomenologies for their cosmological evolution. On the other hand, these couplings could also give rise to natural reheating mechanisms within inflationary models.



\renewcommand{\arraystretch}{1.5}
\begin{table}\centering
\begin{tabular}{ccc||c||c||c||c||c||c||c||c||c||ccc}
\hline 
 & Scalar $\chi$& \multicolumn{11}{c}{} & \multicolumn{2}{c}{Pseudo-scalar $\phi$}\tabularnewline
\cline{2-2} \cline{14-15} 
 & $b_{1}>0$ & \multicolumn{11}{c}{} & $m_{S}^{2}>0$ & $m_{S}^{2}<0$\tabularnewline
\cline{2-15} 
$m_{T}^{2}>0$ & $\chi<\frac{3}{4}m_{T}^{2}$ & \multicolumn{11}{c}{} & $|\phi|<\left|\frac{3m_{S}m_{T}}{4\alpha}\right|$ & {\color{red}Ghost}\tabularnewline
$m_{T}^{2}<0$ & $\chi>\frac{3}{4}m_{T}^{2}$ & \multicolumn{11}{c}{} & {\color{ForestGreen}Healthy} & $|\phi|>\left|\frac{3m_{S}m_{T}}{4\alpha}\right|$\tabularnewline
\hline 
\end{tabular}
\caption{This table summarises the conditions to avoid ghosts for the scalar and the pseudo-scalar field.}
\label{table}
\end{table}

\renewcommand{\arraystretch}{1}

\subsection{The general healthy bi-scalar theory}

For completeness, we will analyse now the theory that propagates simultaneously the scalar and pseudo-scalar fields obtained above. It should be clear that the corresponding theory will be described by the Lagrangian
\be
\Lag=\frac12\mpl^2R+\frac12m_T^2T^2+\frac12m_S^2S^2+b_1R^2+\alpha\mathcal{H}^2.
\ee
We will proceed analogously by introducing auxiliary fields, but we will omit unnecessary details here, which exactly follow the developments of the previous sections. The transformed Lagrangian in the post-Riemannian expansion can be written as
\be
\Lag=\mU(\chi,\phi)+\chi\Rb+\frac12M_T^2(\chi)T^2+\frac12M_S^2(\chi)S^2+\frac43\alpha\phi S_\mu T^\mu-2T^\mu\partial_\mu\chi+2\alpha S^\mu\partial_\mu\phi\,,
\label{eq:bisint}
\ee
where we have defined
\bea
\mU=-\frac{\big(2\chi-\mpl^2\big)^2}{16b_1}-\alpha\phi^2,\quad M_T^2=m_T^2-\frac43\chi\quad\text{and}\quad 
M_S^2=m_S^2+\frac{1}{12}\chi.
\eea
A more useful and compact way of writing the Lagrangian is
\be
\Lag=\mU(\chi,\phi)+\chi\Rb+\frac12\vec{Z}^t\hat{M}\vec{Z}+\vec{Z}^t\cdot\vec{\Phi}
\ee
with $\vec{Z}^t=(T_\mu,S_\mu)$, $\vec{\Phi}^t=(-2\partial_\mu\chi,2\alpha\partial_\mu\phi)$ and
\bea
\hat{M}=
\left(
\begin{array}{cc}
  M_T^2(\chi)&\frac43\alpha\phi  \\
  \frac43\alpha\phi&M_S^2(\chi)  \\
\end{array}
\right).
\eea
The equations for $S^\mu$ and $T^\mu$ can then be written as
\be
\hat{M}\vec{Z}=-\vec{\Phi}\quad\Rightarrow\quad \vec{Z}=-\hat{M}^{-1}\vec{\Phi},
\ee
with the inverse of $\hat{M}$ given by
\bea
\hat{M}^{-1}=\frac{1}{M_S^2(\chi)M_T^2(\chi)-\left(\frac43\alpha\phi\right)^2}
\left(
\begin{array}{cc}
  M_S^2(\chi)&-\frac43\alpha\phi  \\
  -\frac43\alpha\phi&M_T^2(\chi)  \\
\end{array}
\right).
\eea
By inserting this solution into the Lagrangian we finally obtain
\be
\Lag=\mU(\chi,\phi)+\chi\Rb-\frac12\vec{\Phi}^t\hat{M}^{-1}\vec{\Phi}.
\ee
It is then very clear that the theory indeed describes two propagating scalars. We can write out the above compact form of the Lagrangian to make everything more explicit
\be
\Lag=\chi\Rb+6\frac{3M_S^2(\chi)(\partial\chi)^2+3\alpha^2M_T^2(\chi)(\partial\phi)^2-8\alpha^2\phi\partial_\mu\phi\partial^\mu\chi}{(4\alpha\phi)^2-9M_S^2(\chi)M_T^2(\chi)}+\mU(\chi,\phi).
\ee
It is easy to see that this Lagrangian reduces to \eqref{eq:BDPGT} for $\phi=0$ and to \eqref{eq:Holst4} for $\chi=0$ (except for the Einstein-Hilbert term that should be added), as one would expect. The general discussions for the $R^2$ and Holst square theories then also apply to the present case. We see that the scalar $\chi$ features a non-minimal coupling that can be removed by means of the same conformal transformation as before $\gt_{\mu\nu}=\frac{2\chi}{\mpl^2}g_{\mu\nu}$. After performing this transformation to the Einstein frame the Lagrangian reads
\bea
\Lag&=&\frac12\mpl^2\tilde{R}-\left[1-\frac{12M_S^2(\chi)}{(4\alpha\phi)^2-9M_S^2(\chi)M_T^2(\chi)}\right](\partial\chi)^2\nonumber\\
&&+\frac{3\mpl^2}{\chi}\frac{3\alpha^2M_T^2(\chi)(\partial\phi)^2-8\alpha^2\phi\partial_\mu\phi\partial^\mu\chi}{(4\alpha\phi)^2-9M_S^2(\chi)M_T^2(\chi)}+\left(\frac{\mpl^2}{2\chi}\right)^2\mU(\chi,\phi).
\label{eq:biscalar2}
\eea
Again, the conformal transformation will couple $\chi$ directly to matter through the conformal metric, while the pseudo-scalar $\phi$ only couples to the axial fermionic current given by the dual of the corresponding hypermomentum. The same reasoning used to obtain \eqref{eq:Holst5} applies here so this axial coupling eventually generates couplings achievable via the replacement $2\alpha\partial_\mu\phi\rightarrow 2\alpha\partial_\mu\phi+\Delta_\mu$ in  \eqref{eq:biscalar2}. Notice that additional couplings between $\chi$ and fermions will be generated by this mechanism. The resulting Lagrangian \eqref{eq:biscalar2} resembles a two dimensional non-linear sigma model with the following target space metric:
\bea
h_{ab}(\chi,\phi)=\frac{2\mpl^2}{\chi}
\left(
\begin{array}{cc}
\frac{3}{4}+ \frac{1}{\chi}(\hat{M}^{-1})_{11} &\alpha(\hat{M}^{-1})_{12}  \\
  \alpha(\hat{M}^{-1})_{12}&\alpha^2(\hat{M}^{-1})_{22}  \\
\end{array}
\right).
\eea
The resemblance is only formal at this point due to the pseudo-scalar nature of $\phi$ that obstructs its interpretation as a coordinate of the would-be target manifold. The ghost-free conditions are obtained by imposing the positivity of the eigenvalues of this metric, whose expressions are more involved in this case because of the couplings between both scalars. A much simpler condition can be obtained by computing the determinant
\be
\det h_{ab}=\frac{\mpl^4\alpha^2}{\chi^3}\frac{3M_T^2(\chi)+4\chi}{M_S^2(\chi)M_T^2(\chi)-\left(\frac43\alpha\phi\right)^2}
\ee
which must be positive to guarantee ghost-freedom, although this is not a sufficient condition. Moreover, having $\det h_{ab}=0$ will determine the degenerate cases where the phase space is reduced. This happens trivially for $\alpha=0$, that corresponds to the pure $R^2$ theory. The pure Holst square limit is more delicate to obtain because the conformal transformation becomes singular for $\chi=0$. We will not explore further the general bi-scalar theory here, but it should be clear that such theories will contain a much richer structure owed to its enlarged phase space.

We will end our discussion of the bi-scalar theories by explicitly showing how our results can be straightforwardly extended to theories described by a general function of $R$ and $\mathcal{H}$. Let us then consider the following Lagrangian:
\be
\Lag=F(R,\mH,T,S,q),
\ee
where $F$ is some arbitrary scalar function. Also, for the sake of generality, we have allowed an arbitrary dependence on the torsion components as well. The Lagrangian can be recasted in the form
\be
\Lag=F(\tilde{\chi},\tilde{\phi},T,S,q)+\chi \big(R-\tilde{\chi}\big)+\phi\big(\mH-\tilde{\phi}\big)
\ee
where we have introduced a set of auxiliary fields. The equations for $\tilde{\chi}$ and $\tilde{\phi}$ allow to express these fields in terms of the rest of fields. We can then write
\bea
\Lag=\mU(\chi,\phi,T,S,q)+&\chi \left(\Rb+\frac{1}{24}S^2-\frac{2}{3}T^2+2\nablab_\mu T^\mu+\frac12q_{\mu\nu\rho}q^{\mu\nu\rho}\right)\nonumber\\
&+\phi\left(\frac23 S_\mu T^\mu-\nablab_\mu S^\mu+\frac12\epsilon_{\alpha\beta\mu\nu}q_\lambda{}^{\alpha\beta}q^{\lambda\mu\nu}\right),
\eea
where the potential $\mU$ already includes the effects of integrating out $\tilde{\chi}$ and $\tilde{\phi}$. Again, we see that the pure tensor sector only enters as an auxiliary field so we can also integrate it out to finally write the Lagrangian as
\be
\Lag=\tilde{\mU}(\chi,\phi,T,S)+\chi\Rb-2T^\mu\partial_\mu\chi+S^\mu\partial_\mu\phi
\ee
with $\tilde{\mU}$ containing all the terms without derivatives. This Lagrangian resembles \eqref{eq:bisint} with the only difference that the non-derivative terms are different. We can thus proceed analogously by integrating out the vector sector $T_\mu$ and $S_\mu$ by solving their equations of motion
\bea
\frac{\partial\tilde{\mU}}{\partial T^\mu}-2\partial_\mu\chi=0\, ,\\
\frac{\partial\tilde{\mU}}{\partial S^\mu}+\partial_\mu\phi=0\, ,
\eea
that will give $T_\mu=T_\mu(\chi,\phi,\partial\chi,\partial\phi)$ and $S_\mu=S_\mu(\chi,\phi,\partial\chi,\partial\phi)$. By plugging these solutions back in the Lagrangian we finally arrive at the explicit bi-scalar theory, but now with more involved interactions that will depend on the specific function describing the Lagrangian. If we include couplings to fermions, the same trick as before can be used to take it into account.

\subsection{Adding dimension 4 operators}

We have seen how to restrict the parameters in order to remove the ghosts of the quadratic PGTs while having an additional dynamical scalar. We will now discuss how to tame the ghosts by extending the Lagrangian in a suitable manner. For that, it is worthwhile to notice that the constructed quadratic theories contain up to dimension 4 operators corresponding to the curvature squared terms. It would then seem natural to include all the operators up to that dimensionality. For instance, since the Riemann squared terms generate quartic interactions for the torsion, there seems not to be a reason why they should not be included from the onset of the construction of the theory. If we do allow for all the operators up to dimension four, there is a whole bunch of additional torsion terms that we could add. In particular, we can include the operators $\mT_{\mu\nu} \mT^{\mu\nu}$ and $\mS_{\mu\nu} \mS^{\mu\nu}$ with arbitrary coefficients. In the presence of these additional terms, it is trivial to see that the unavoidable ghostly nature of the vector sector concluded above by removing dangerous non-minimal couplings is resolved. Furthermore, since these are just standard Maxwell terms, they will tackle the ghosts issue without introducing new potentially pathological interactions for the vector sector and affecting the pure tensor sector.

Once the presence of arbitrary dimension 4 operators is allowed, we can also include other phenomenologically interesting interactions. In particular, we can add non-minimal couplings that do not spoil the stabilisation achieved by including the aforementioned Maxwell terms. For instance, we can introduce interactions that mix the curvature and the torsion. Generically, these interactions will be pathological. There is however a class of operators that gives rise to non-pathological non-minimal couplings for the vector sector. That is the case of $G_{\mu\nu} T^\mu T^\nu$ that generates the following couplings in the post-Riemannian expansion
\be
\Lag\supset \Gb _{\mu\nu} T^\mu T^\nu-T^2\nablab_\mu T^\mu+\frac13 T^4-\frac{1}{144}S^2 T^2-\frac{1}{72} (S_\mu T^\mu)^2
\ee
that includes the non-minimal coupling to the Einstein tensor and a vector-Galileon term for the vector trace. When turning on the tensor piece however some other worrisome terms will also enter which could potentially jeopardise the stability of the vector sector.

\section{Discussion}
In this note we have shown that imposing ghost-freedom in the quadratic Poincar\'e gauge theory around an arbitrary background generically leads to a non-dynamical torsion sector so that, after integrating it out, we have nothing but GR, with perhaps some effects in the fermion interactions. This result illustrates once more the delicate nature of gravity and the difficulty to construct consistent modifications of GR. The origin of the instabilities has been clearly traced to the presence of quadratic curvature invariants in the Lagrangian that generate ghostly non-minimal couplings and non-gauge-invariant derivative interactions for the vector sector. The pathological nature of higher order curvature terms in the Lagrangian is in fact a common problem within general metric-affine theories \cite{BeltranJimenez:2019acz}. Even if we restrict to curvature free geometries, general teleparallel theories are generically plagued by the same pathologies \cite{BeltranJimenez:2017tkd,Conroy:2017yln,BeltranJimenez:2018vdo,Koivisto:2018loq,Jimenez:2019tkx,Jimenez:2019ghw}. In most of these theories, the shortcomings for their stability ultimately resides in the presence of additional fields which generically exhibit non-minimal couplings that are at the heart of the harmful ghostly modes. It frequently occurs that these pathologies do not show up in perturbative analysis around highly symmetric backgrounds  where some modes may even disappear, thus giving a false impression of stability. However, these latent modes are even more virulent because the lack of dynamics around the considered backgrounds typically signals an additional pathology in the form of strong couplings.

For the quadratic PGTs analysed in this work, we have seen that the torsion trace does not introduce any pathologies and this has been proven to be the case also in higher dimensions, where healthy non-minimal couplings arise. The source of the problems has been identified to reside in the axial sector. We have seen that getting rid of its ghostly interactions requires a condition on the parameters that forces either $T_\mu$ or $S_\mu$ to be a ghost so we are eventually forced to impose them to be non-dynamical to avoid ghosts. It is important to emphasise that this strong result cannot be obtained from a perturbative analysis around Minkowski because the ghost actually originates from problematic interactions of the axial sector that trivialise at linear order around Minkowski. 

After demonstrating the pathological nature of general quadratic PGTs, we have discussed how to construct healthy theories. We have explicitly worked out the theory whose quadratic sector in curvatures reduces to $R^2$. In this particular theory, the only non-trivial part of the torsion is the vector trace which in turn reduces to a scalar field and the resulting Lagrangian is nothing but a generalised Brans-Dicke theory. Also, by going to the Einstein frame we have revealed the dynamical nature of the scalar as a consequence of the breaking of a conformal symmetry induced by the mass terms of the torsion. Furthermore, the Einstein frame description of the theory makes the dynamics of the scalar apparent and permits a more direct comparison with known studies such as those dealing with cosmologies or black hole solutions with scalar fields. For instance, the conformal coupling of the scalar to matter makes the case for the presence of screening mechanism of the chameleon \cite{Khoury:2003aq} or symmetron \cite{Hinterbichler:2010es} type.

On the other hand, we have shown that under a specific choice of the parameters, it is also possible to have a pseudo-scalar propagating in a stable manner. This particular theory has been shown to be related to the Holst formulation of GR where the Barbero-Immirzi parameter is promoted to a pseudo-scalar field which is precisely the propagating $0^-$ already identified in e.g. \cite{Hecht:1996np,Yo:1999ex}. After integrating out all the auxiliary fields in the theory, we have obtained a final expression for the Lagrangian with the explicit form of the potential and from which the stability conditions are easily obtained. As we have discussed, these theories offer compelling scenarios for dark matter as coherent oscillations of the pseudo-scalar field and accelerated cosmological solutions with relevance for dark energy and inflation. Moreover, although the pseudo-scalar does not couple through a conformal factor to matter, we have seen how the coupling of the axial vector to fermions predicts a natural derivative coupling between the pseudo-scalar and fermions, which could lead to interesting cosmological phenomenology worth exploring. 

Finally, we have obtained the explicit bi-scalar formulation of the theory containing both $R^2$ and Holst square terms and we have also argued how the addition of other dimension 4 operators allow to trivially stabilise the vector sector of the quadratic PGTs. This is a natural possibility because the theories already contain operators of this dimensionality. Moreover, we have also discussed how  allowing for arbitrary dimension 4 operators opens the prospects for constructing   ghost-free non-minimal couplings and Galileon-like interactions within the framework of PGTs.

In summary, we have confirmed existing results in the literature concerning the stability of PGTs by following an alternative, perhaps more direct, approach that gives a complementary understanding of these theories. Furthermore, we have provided a more insightful description of the known stable theories featuring scalar modes by explicitly constructing their effective Lagrangians and showing their relation with standard scalar-tensor theories and the Holst formulation of GR. In view of our findings, the solutions of these theories (cosmological, black holes, wormholes, etc.) can be better contextualised and easier to interpret in terms of the scalar interactions. By reverse engineering, the alternative descriptions presented in this work also allow to find new solutions for PGTs. Hopefully, these equivalent formulations pave the way for a more systematic, exhaustive and physically appealing exploration of the solutions and phenomenological applications of PGTs.

\begin{acknowledgements}
We would like to thank Marc Mars for useful discussions and \'Alvaro de la Cruz, Alejandro Jim\'enez-Cano and Yuri Obukhov for their comments. We would also like to thank the authors of \cite{Vasilev:2017twr} for discussing their results with us. JBJ acknowledges support from the  {\textit{ Atracci\'on del Talento Cient\'ifico en Salamanca}} programme and the MINECO's projects FIS2014-52837-P and FIS2016-78859-P (AEI/FEDER).  This article is based upon work from COST Action CA15117, supported by COST (European Cooperation in Science and Technology). FJMT acknowledges financial support from UCT Launching Grants Programme and NRF Grants No. 99077 2016-2018, Ref. No. CSUR150628121624, 110966 Ref. No. BS170509230233, and the NRF IPRR, Ref. No. IFR170131220846. FJMT also acknowledges support from the Erasmus+ KA107 Alliance4Universities programme and from the Van Swinderen Institute at the University of Groningen. FJMT thanks the hospitality of the University of Salamanca (Spain) during the development of the manuscript.
\end{acknowledgements}
\bibliographystyle{JHEP}
\bibliography{References}

\end{document}